\documentclass[twocolumn]{aastex701} 

\usepackage{natbib,amsmath}
\usepackage{graphicx}
\usepackage{amssymb}
\usepackage{float}
\usepackage{makeidx}
\usepackage{amsmath}
\usepackage{rotating}
\usepackage{longtable}
\usepackage{subfigure}
\usepackage{morefloats}
\usepackage{captcont}
\usepackage{xcolor} %text color
\usepackage{chngcntr}
\usepackage{wrapfig}
\usepackage{multirow}
\usepackage[multiple]{footmisc} 
\usepackage{url}
\usepackage{enumitem}

\graphicspath { {./Figures/}}

\def\N#1{{N({\rm #1})}}
\newcommand{\etal}{et~al.\/}

\usepackage{tabularx}

\newcommand\clearrow{\global\let\rowmac\relax}
\clearrow

\shortauthors{Bish \etal}
\shorttitle{LightCube 3D ISRF}

\begin{document}
%%%%%%%%%%%%%%%%%%%%%%%%%
\title{\textit{LightCube}: A Parsec-Resolution 3D Model of the Local Far-Ultraviolet Interstellar Radiation Field}

\author[0000-0002-7483-8688]{Hannah V. Bish}
\affiliation{Space Telescope Science Institute, 3700 San Martin Drive, Baltimore, MD 21218, USA}
\email{hbish@stsci.edu}

\author[0000-0003-4797-7030]{J. E. G. Peek}
\affiliation{Space Telescope Science Institute, 3700 San Martin Drive, Baltimore, MD 21218, USA}
\affiliation{Department of Physics \& Astronomy, Johns Hopkins University, 3400 North Charles Street, Baltimore, MD 21218, USA}
\email{jegpeek@stsci.edu}

\author[0000-0002-2250-730X]{Catherine Zucker}
\affiliation{Center for Astrophysics $|$ Harvard \& Smithsonian, 60 Garden Street, Cambridge, MA 02138, USA}
\email{catherine.zucker@cfa.harvard.edu}

\author[0000-0001-5340-6774]{Karl D. Gordon}
\affiliation{Space Telescope Science Institute, 3700 San Martin Drive, Baltimore, MD 21218, USA}
\affiliation{Department of Physics and Astronomy, Universiteit Gent, Proeftuinstraat 86 N3, B-9000 Ghent, Belgium}
\email[hide]{kgordon@stsci.edu} 

\author[0000-0002-7743-8129]{Claire E. Murray}
\affiliation{Space Telescope Science Institute, 3700 San Martin Drive, Baltimore, MD 21218, USA}
\email{cmurray1@stsci.edu}

\author[0000-0002-7633-3376]{Susan E. Clark}
\affiliation{Department of Physics, Stanford University, Stanford, CA 94305, USA}
\email{seclark1@stanford.edu}
\affiliation{Kavli Institute for Particle Astrophysics \& Cosmology, P.O. Box 2450, Stanford University, Stanford, CA 94305, USA}

\author[0009-0000-1042-5772]{Mathieu Willems}
\affiliation{Space Telescope Science Institute, 3700 San Martin Drive, Baltimore, MD 21218, USA}
\affiliation{Department of Physics and Astronomy, Universiteit Gent, Proeftuinstraat 86 N3, B-9000 Ghent, Belgium}
\email{mwillems@stsci.edu}

\author[0000-0002-3131-7372]{Erika Hamden}
\affiliation{Steward Observatory, University of Arizona, Tucson, AZ 85719, USA}
\email{hamden@arizona.edu}

% #####################################
% ############# ABSTRACT ##############
% #####################################
\begin{abstract}

We present a three-dimensional model of the local interstellar radiation field (ISRF) in the ultraviolet (UV). Using UV flux measurements from the TD1 catalog and stellar distances from \emph{Gaia} and Hipparcos, we construct a catalog of stars that we expect to dominate the UV flux in the nearby Galaxy. We use the radiative transfer code DIRTY to model the propagation of photons from these stars through a 3D dust map, including the effects of scattering and absorption. The result is \textit{LightCube}, a model of the ultraviolet ISRF out to 1.25 kpc from the Sun, with a maximum linear resolution of 1 pc. We model the ISRF in the four TD1 bands (1565 $\rm\AA$, 1965 $\rm\AA$, 2365 $\rm\AA$, and 2740 $\rm\AA$), as well as a single value for the full FUV range, and calculate the ISRF at the Sun to be 5.65$\times$10$^{-14}$ erg cm$^{-3}$ from 912 $\rm\AA$ to 2000 $\rm\AA$. The modeled ISRF is quite variable, with more than an order of magnitude variation seen in dense regions, and about half that in lower density regions. By comparing \textit{LightCube} to an estimate of the 3D distribution of total-to-selective extinction ratio, $R_V$, we find a positive correlation between UV flux and $R_V$ in regions of low UV radiation.

\end{abstract}


%%%%%%%%%%%%%%%%%%%%%%%%%%%%%%%%% %%%%%%%%%%%% %%%%%%%%%%%%%%%%%%%%%%%%%%%%%%%%%
%%%%%%%%%%%%%%%%%%%%%%%%%%%%%%%%% INTRODUCTION %%%%%%%%%%%%%%%%%%%%%%%%%%%%%%%%%
%%%%%%%%%%%%%%%%%%%%%%%%%%%%%%%%% %%%%%%%%%%%% %%%%%%%%%%%%%%%%%%%%%%%%%%%%%%%%%
\section{Introduction}
\label{sec:intro}

The interstellar radiation field (ISRF) – the photon field produced by stars and interstellar matter that permeates interstellar space – is thought to be responsible for setting many of the properties of diffuse matter in galaxies \citep{Wolfire1995}. Despite its crucial role in governing many processes throughout the interstellar medium (ISM), including dust destruction \citep{Allain96}, ISM heating \citep{Wolfire1995}, molecule destruction \citep{DB96}, and birth cloud destruction \citep{K14}, surprisingly little is understood about the ISRF's distribution in 3D space. 

Early foundational work on the ISRF was done by \cite{Habing1968}, who calculated the average energy density of the ISRF at the Sun by counting early type stars and modeling their radiation field. This work established the Habing field, $G_0$, which is often assumed to hold for all locations in the Solar neighborhood. Later \cite{Mathis1983} provided a more comprehensive spectrum at the solar neighborhood, from UV to infrared, and modeled its interactions with dust grains. This work also extended the Habing fixed field to a one-dimensional field, with a varying ISRF as a function of Galactocentric radius. The study of the local ISRF has continued to the present day, including the work of \citet{Bianchi2024} who spectroscopically modeled \textit{Gaia} and \textit{Hipparcos} to create a more precise ISRF across the UV and optical bands. In order to model the gamma ray continuum, \cite{Strong00} developed a two-dimensional ISRF, as a function of both galactocentric radius and height above the disk. This allowed them to model the inverse Compton scattering process of cosmic rays off of the anisotropic ISRF. This cosmic ray analysis was also the major driver for the first 3D models of the ISRF by \cite{Porter08}. \cite{Popescu17} added more sophisticated radiative transfer techniques to this approach, as well as large-scale anisotropic structures like the bar and spiral arms.

These approaches are powerful for modeling relatively smooth, low-resolution observations like the gamma-ray sky, but cannot capture the detailed, highly variable structure of the ISM in the Solar neighborhood. Many observational tracers that depend on the ISRF, such as absorption lines for gas ionization state \citep{JT01}, mid- and far-IR emission for dust temperature \citep{Bernard10}, and radio emission for molecular state \citep{HD15}, are sensitive to fine scale variation in the ISM and the ISRF. As a consequence, our current lack of detailed 3D information about the ISRF means we do not have a clear understanding of the role it plays in these processes and how to interpret observations.

Many sources contribute to the ISRF across the electromagnetic spectrum, but the ultraviolet (UV) regime is the most crucial to model for several reasons. This energy range is important for photoexcitation of molecules \citep{black1987, sternberg1989} and ionization of metals \citep{heays2017}. It can also eject electrons from dust grains, making it a dominant heating mechanism of the bulk of the ISM \citep{Draine78, bakes1994, weingartner2001}. Thus, the ISRF in large part sets the thermal balance of cold and warm neutral clouds \citep{Wolfire2003}. Additionally, the UV is hugely influential in shaping emission in the infrared (IR), because UV radiation absorbed by dust is then re-radiated in the IR \citep{BolleaCavaliere1976,daCunha2008}. %  \citep{Bialy2020}

Unlike the ISRF at other wavelengths, the UV ISRF is dominated by radiation from young, high-mass stars, which are highly clustered in space and last for relatively short periods of time. UV photons are also preferentially scattered off of dust grains and interact strongly with hydrogen gas below the Lyman limit. While diffuse clouds are transparent to most IR radiation, shorter wavelengths in the UV are much more likely to be absorbed or scattered. Together, these factors make the UV ISRF particularly challenging to model because of its variability, and despite being a key driver of ISM physics it is therefore often assumed to be constant across space. 

The \cite{Habing1968} estimate of the integrated energy density in the range \mbox{6-13.6 eV} remains a benchmark reference value for ISRF measurements. A `Habing', annotated as $G_0$ with a value of $1.6\times10^{-3}$ erg cm$^{-2}$ s$^{-1}$, is still a standard unit of ISRF measurement. \cite{Draine78} collated a number of measurements to arrive at a ``Standard" ISRF of, roughly, 1.7 times higher than $G_0$. We will refer to this as the Standard ISRF throughout the remainder of this work. We note that while this is by no means the most up to date radiation field, it is both a well known benchmark, and refers only to the UV field we are able to probe in this work.

Two recent advances have enabled detailed modeling of the 3D UV ISRF. First, \emph{Gaia} has brought about a revolution in astrometry \citep{gaiamission}, providing a much more complete 3D map of the stars in our solar neighborhood and allowing us to obtain precise distances to the sources of UV radiation. Second, dust extinction maps have improved dramatically in recent years \citep{edenhofer2024}, and we now have extremely detailed 3D representations of the dust that is responsible for absorbing and scattering the UV light in the nearby ISM. 

Here we present a new model that uses a UV-bright source catalog, recent high-resolution dust mapping, and 3D dust radiative transfer to calculate a detailed 3D map of the local UV ISRF. We demonstrate that the ISRF varies by orders of magnitude across the volume. We provide access to the data in 3D so it can be compared directly to observations of the sky. We hope this will serve as a foundational tool for future studies on the interplay between radiation and matter in our Galaxy.

The remainder of this paper is structured as follows. In \S\ref{sec:methods} we outline our methods, including the UV source catalog processing and radiative transfer calculations. In \S\ref{sec:results} we present the resulting \textit{LightCube} model and detail basic properties. We discuss the implications of these results and the correlation between UV flux and $R_V$ in \S\ref{sec:discussion}, and provide a summary of this work in \S\ref{sec:summary}.

%%%%%%%%%%%%%%%%%%%%%%%%%%%%%%%%% %%%%%%%%%%%% %%%%%%%%%%%%%%%%%%%%%%%%%%%%%%%%%
%%%%%%%%%%%%%%%%%%%%%%%%%%%%%%%%%%%% METHODS %%%%%%%%%%%%%%%%%%%%%%%%%%%%%%%%%%%
%%%%%%%%%%%%%%%%%%%%%%%%%%%%%%%%% %%%%%%%%%%%% %%%%%%%%%%%%%%%%%%%%%%%%%%%%%%%%%
\section{Methods}
\label{sec:methods}

\begin{figure*}%[h!]
    \includegraphics[width=0.9\linewidth]{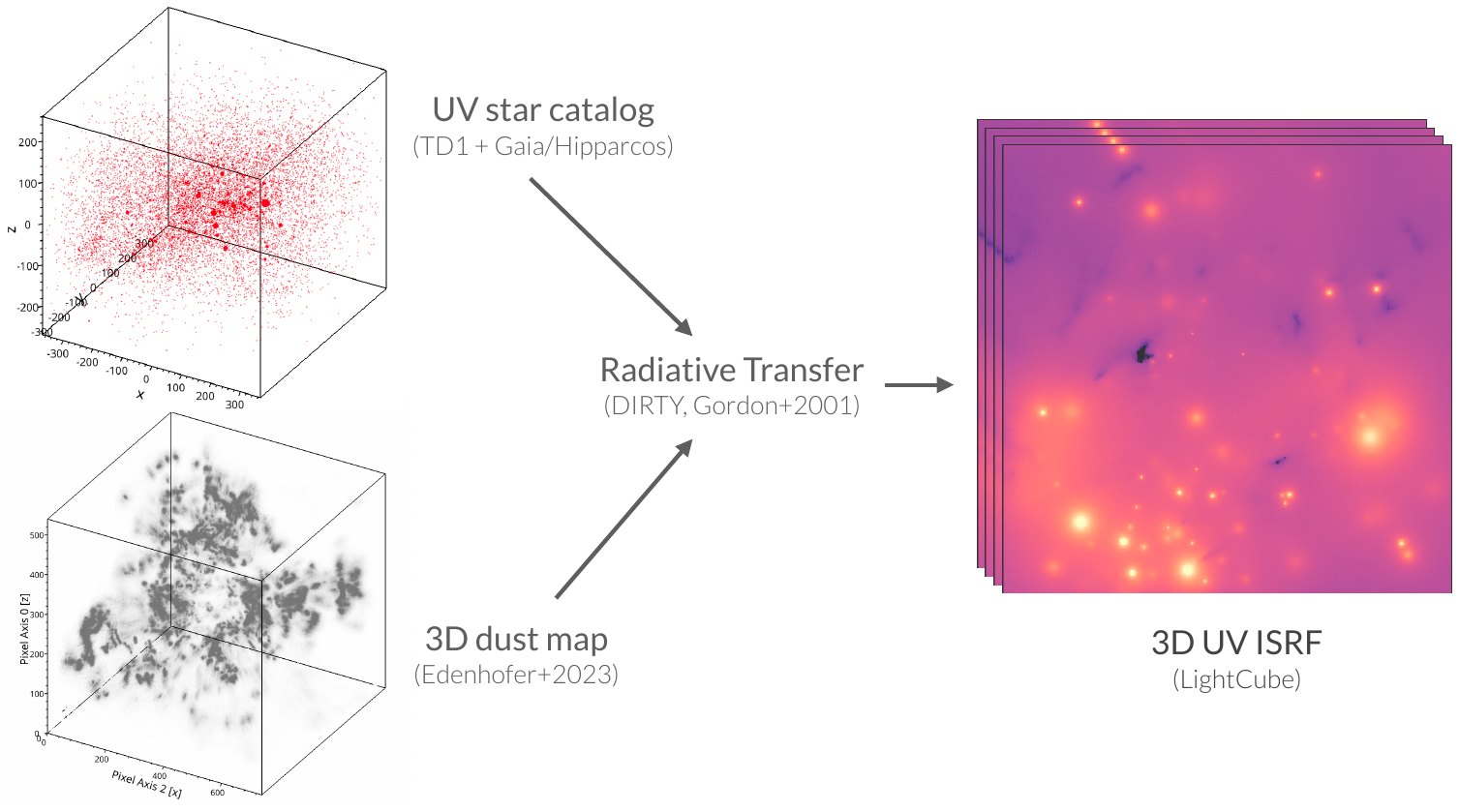}
    \caption{\textsc{Building the LightCube ISRF model.} We cross-match the TD1 catalog of UV source fluxes \citep{TD11995} with distances from Gaia \citep{gaiaDR3summary} and Hipparcos \citep{hipparcos1997}, allowing us to place these sources within a 3D dust map of the local ISM \citep{edenhofer2024} We use DIRTY, a radiative transfer code \citep{Gordon2001, Misselt2001}, to trace UV photons from these sources as they move through the cube and are absorbed or scattered by the dust. The result is LightCube, a 3D model of the radiation field density in the local ISM.
    \label{fig:flowchart}}
\end{figure*}

The combination of stellar astrometry and high-resolution dust maps provides the ingredients for building our new high-resolution, 3D model of the UV ISRF. With \textit{Gaia}’s precise positions and luminosities of nearby stars, we can pinpoint the sources of UV radiation in 3D space. We place those stars within a high-resolution 3D dust map and use a radiative transfer code to trace photons through space as they are emitted from the stars and are scattered or absorbed by dust. The result is a data-driven, three-dimensional model of the radiation field with resolution as high as 1 parsec that extends out to 1.25 kpc from the Sun, which we dub \textit{LightCube}. 
Below we detail the process we follow to assemble \textit{LightCube} (see Figure \ref{fig:flowchart}). 

In order to obtain UV fluxes for these stars we turn to the TD1 Stellar Ultraviolet Fluxes Catalog \citep{Thompson1978,TD11995}. This catalog is the product of an all sky survey by the Belgian/UK Ultraviolet Sky Survey Telescope (S2/68) aboard the USRO TD1 satellite conducted in 1972. This is the most recent all-sky catalog of UV-bright sources, and covers approximately 31,000 sources in four UV bands centered on 1565 \AA, 1965 \AA, 2365 \AA, and 2740 \AA. These data In addition to fluxes, we also obtain accurate 3D coordinates for all of the stars in TD1 by crossmatching them with Gaia \citep{gaiaDR2summary,gaiaDR3summary}. For the brightest stars, which Gaia cannot observe, we perform a crossmatch with Hipparcos instead \citep{hipparcos1997}. The crossmatching process is described in more detail in \S\ref{subsec:crossmatching}. 

\begin{figure}[h!]
    \includegraphics[width=\linewidth]{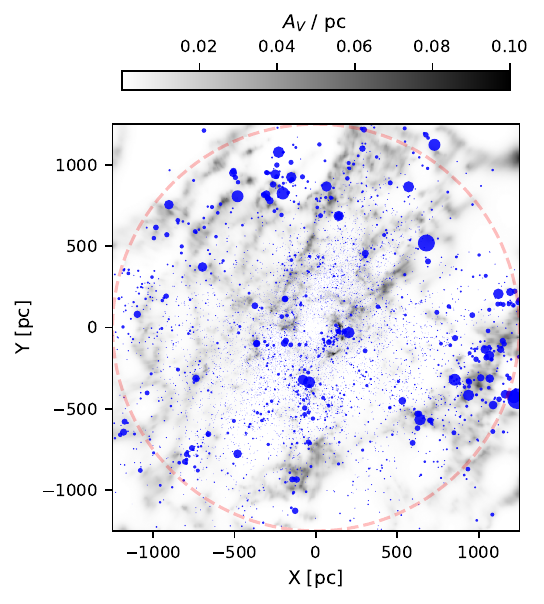}
    \caption{\textsc{Map of stars and dust in the cube.} Top-down view of the \cite{edenhofer2024} dust map with UV bright stars. The dust map is shown averaged over the Galactic midplane at -100 pc $<$ z $<$ 100 pc, with the Sun at the origin. Blue points show all the stars included in the model, with the size of the point indicating the area over which the star's radiation is greater than the Standard ISRF. The dashed red line represents the boundary of the reliable \cite{edenhofer2024} map.
    \label{fig:edenhofer-map}}
\end{figure}

We use the \citet{edenhofer2024} 3D dust map, which provides the differential dust extinction out to 1.25 kpc in all directions (see Figure \ref{fig:edenhofer-map}). The map was constructed using distance and extinction estimates for roughly 54 million stars derived from Gaia BP/RP spectra, supplemented by infrared photometry from 2MASS and unWISE \citep{Zhang2023}. The spatial distribution of differential dust extinction was inferred via a Gaussian process prior, enforced using iterative charted refinement \citep{edenhofer2022}, which mitigates the fingers-of-god artifact common in previous reconstructions. Posterior inference was performed using metric Gaussian variational inference \citep{knollmuller2019}, yielding 12 posterior samples over a grid of 516 logarithmically spaced HEALPix shells at $\rm N_\mathrm{side} = 256$. The distance resolution ranges from roughly one parsec near the Sun to approximately eight parsecs at the edges of the map. We utilize the released mean map (averaged over the 12 posterior samples) in this work.

We combine the coordinate and luminosity information from UV-bright stars with the high-resolution dust maps by using both as inputs to a radiative transfer code, DIRTY \citep{Gordon2001,Misselt2001,Law2018}. DIRTY takes as input the location and luminosity of photon sources within the dust map, performs the full 3D dust radiative transfer \citep{Steinacker2013}, and computes the resulting radiation density throughout the map. We include the most luminous 17,000 stars in the TD1 catalog that fall within the dust map volume; stars less luminous than these produce a radiation density of less than $G_0$ at a distance of 1 pc (the smallest voxel size in the cube), and therefore do not meaningfully impact the radiation density beyond the voxel in which they sit.

%%%%%%%%%%%%%%%%%%%%%%%%%%%%%%%%%%%%%%%%%%%%%%%%%%%%%%%%%%%%%%%%%%%%%%%%%%%%%%%%
\subsection{Crossmatching \& Sample Selection}
\label{subsec:crossmatching}
%%%%%%%%%%%%%%%%%%%%%%%%%%%%%%%%%%%%%%%%%%%%%%%%%%%%%%%%%%%%%%%%%%%%%%%%%%%%%%%%

Our sample begins with the TD1 catalog \citep{TD11995}, which has positions and UV fluxes for 31,215 nearby stars. In our initial analysis we determined that the coordinates listed in TD1 are not sufficiently precise for accurate crossmatching using coordinates alone (see Appendix \ref{app:dataquality}), and thus we began our crossmatching process based on the HD numbers listed in TD1. We used \textit{Simbad} \citep{wenger2000} for our first pass to determine correct coordinates. 369 objects listed in TD1 did not have HD numbers to match to \textit{Simbad}. Any objects that had a different paired TD1 ID and HD number listed in \textit{Simbad} versus the TD1 catalog were hand inspected to determine which match was correct. In practice, this was typically because binary stars were listed as two objects in \textit{Simbad} and one object in TD1. Any stars that didn’t have a HD number listed were matched to Simbad using the TD1 ID. Following this process, a few hundred objects remained with no match. We manually searched for matches to all of these missing stars with $V$-band magnitude less than 9. Ultimately there was a \textit{Simbad} match for all stars with $V<9$, and 9 duplicate matches.

After this initial cross-match with \textit{Simbad}, we also matched these sources to the \emph{Gaia} DR3 \citep{gaiamission, gaiaDR2summary, gaiaDR3summary} and Hipparcos \citep{hipparcos1997} catalogs using the \emph{Gaia} archive \citep{GaiaArchive}. Coordinates were transformed to the appropriate epoch for crossmatching.
Of 31,215 sources in the TD1 catalog, 30,737 had a match in \emph{Gaia}, with 394 duplicates having multiple matches. 23,470 had a match in Hipparcos with 27 duplicates. 244 had no match in either. 

Of 421 TD1 sources with a duplicate match, 178 were binary stars that were listed as one source in TD1 and two sources in other catalogs. The UV flux from TD1 generally reflects the combined flux from both stars in a binary system, so if a TD1 source had duplicate matches and was listed as a binary in \textit{Simbad}, then we chose the individual star match with the smallest fractional parallax error in order to obtain the most accurate distance measurement for the source. For the remaining 243 duplicates, we discarded any matches where the difference between the V band and \emph{Gaia} BP band magnitudes (V-BP) was more than 1. Those cases are likely erroneous matches because those filters have significant overlap and should be similar in magnitude. Finally, if there were still duplicate matches for a non-binary star, or any star had a match in both \emph{Gaia} and Hipparcos, we chose the one with the smallest fractional parallax error. If a source had neither, we used the parallax listed in \textit{Simbad} (if any). 

For stars with $V<9$ that had no match at all using the above steps, we manually searched \textit{Simbad} for possible matches. 
A number of these were also binary stars which didn’t match with Simbad because the “parent” listing for the binary star was a separate entry from the individual stars in the system. We allowed a maximum parallax/parallax\_error of 5 for inclusion in \textit{LightCube}.
Stars with negative fluxes or parallaxes, or no distance information available, were not included in the cube.

For stars that met all these criteria, we selected stars within the 1.25 kpc radius of the \citet{edenhofer2024} dust map volume that had at least one reported luminosity from TD1 in one of the four TD1 bands, a total of 26,519 stars. Of these, 3,311 are missing luminosities in at least one band. To restore the missing luminosities we turn to the k- nearest neighbors (kNN) machine learning method. We train the kNN regressor, as implemented in scikit-learn, on the luminosities of all stars that have an observed flux signal-to-noise of at least 5 in all bands. We then impute the results for all missing luminosities, using k=15, which balances minimizing noise with maximizing the domain over which we can impute the results.

Figure \ref{fig:edenhofer-map} shows the sample of stars included in the model overlaid on the dust map. In the final sample, 50\% of the total flux in the cube is generated by the 158 most luminous stars, 90\% of the flux by the most luminous 2,474 stars, and 99\% of the flux by the most luminous 8,906 stars. We include the 17,000 most luminous stars within the cube in our sample, which generate 99.96\% of the total measured flux. All the remaining stars produce a radiation density of less than $G_0$ at a distance of 1 pc, and are therefore not luminous enough to dominate the radiation field at the edge of the voxel in which they sit. This cutoff provides a computationally tractable sample that is both complete and self consistent, without running into the TD1 sensitivity limit.

\begin{figure}[h!]
    \includegraphics[width=\linewidth]{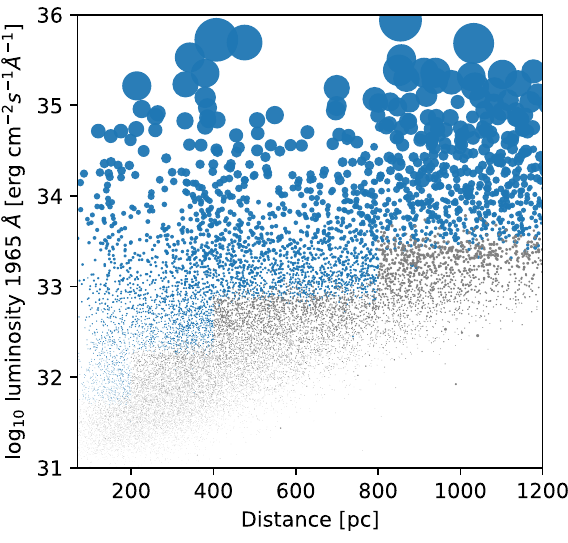}
    \caption{\textsc{Luminosity of stars in the cube.} The absolute magnitude of input stars as a function of distance. Points are scaled such that their radius extends to the distance at which the star would produce the Standard ISRF. While our catalog is not complete to the faintest stars across the \textit{LightCube} volume, the fainter stars do not significantly illuminate much of the volume of space. Stars whose luminosity is low enough that they produce less than the Standard ISRF at the boundary of a voxel where they are centered are shown in gray, and excluded from our model. The step-like boundary between blue and grey points is due to the varying voxel sizes in the grid, which become larger with distance from the Sun.
    \label{fig:completeness}}
\end{figure}

\subsubsection{Extinction Correction}
\label{subsubsec:extinction}
%%%%%%%%%%%%%%%%%%%%%%%%%%%%%%%%%%%%%%%%%%%%%%%%%%%%%%%%%%%%%%%%%%%%%%%%%%%%%%%%

The radiative transfer calculation requires us to know the intrinsic UV luminosity of all sources, so we must correct the observed TD1 fluxes for extinction and calculate the corresponding luminosity using the distances we obtained as described in \S\ref{subsec:crossmatching}. To do this, we first use the \texttt{dustmaps} package \citep{dustmaps2018} to calculate the extinction $A_V$ along the line of sight to the star in the Edenhofer dust map, given the star's coordinates. We then use the \texttt{dust\_extinction} package \citep{dustextinction2024} to convert this extinction value in the $V$ band to corresponding values for the four TD1 bands (with central wavelengths of 1565 \AA, 1965 \AA, 2365 \AA, and 2740 \AA), assuming the extinction model from \citet{Gordon2023} (which in turn used substantial input from \citet{Gordon2009}, \citet{Fitzpatrick2019}, \citet{Gordon2021}, and \citet{Decleir2022}). Finally, we use the \texttt{extinction} package \citep{extinction_barbary_2016} to correct the TD1 fluxes for extinction in all four bands to find the extinction-corrected apparent magnitude, which can then be used along with distance to the star to calculate its luminosity.

\subsubsection{Special Handling of Bright Stars}
\label{subsubsec:missing-fluxes}
%%%%%%%%%%%%%%%%%%%%%%%%%%%%%%%%%%%%%%%%%%%%%%%%%%%%%%%%%%%%%%%%%%%%%%%%%%%%%%%%

Some of the brightest, closest stars may saturate TD1 or have significant nebulosity, either of which could break the photometry methods used to construct the TD-1 catalog; thus we look to other catalogs to ensure we are including such sources. \cite{McCallum2025} published a similar map to \textit{LightCube}, but focused only on the ionizing flux from O stars in the extreme UV. They have compiled and published all the O stars known in the \cite{edenhofer2024} volume, which we check against our catalog. 52 of these stars were matched to stars in TD1, and 35 of them had no counterpart in the catalog. 

Many of the stars listed in \cite{McCallum2025} are binary or multiple star systems, and two are open clusters. In these cases we included each O star in the multiple system separately. For some multiple systems, no information was available for the individual stars, in which case we included the entire multiple system as one `star'. 
In the case of Omicron Persei (aka TD1 2447), which sits near the edge of a dense molecular cloud, the extinction value taken from the \citet{edenhofer2024} dust cube led to predicted luminosities that were approximately five times greater than the next most luminous star; we interpret this as unphysical and instead adopt the extinction value of $E(B-V)=0.32$ mag reported in \citet{Snow1975}. This issue may occur when a star sits near the edge of a dense molecular cloud, as Omicron Persei does, because a small error in distance measurement can lead to a large difference in the extinction value calculated (as described in \S\ref{subsubsec:extinction}). 

For the luminous O stars that were not in TD1 and therefore did not have corresponding UV fluxes available, we used the PHOENIX models \citep{Hauschildt1997} provided in the \texttt{synphot} package \citep{synphot2018} to find the predicted UV fluxes based on the extinction-corrected $V$-band magnitudes, $T_{eff}$, $log(g)$, and metallicity listed in \textit{SIMBAD}. In a small number of cases $V$-band magnitudes were not available, in which case we used the absolute $V$-band magnitude predicted by \cite{Martins2005} based on the star's spectral type. Most stars did not have metallicities available, and in the absence of metallicity information we assumed default values of Fe/H$=$0. If $log(g)$ values were missing we used a default value of 4. If stellar properties fell outside the range covered by the PHOENIX model grids, we used the closest available value in the grid. For the very bright and complex star system $\gamma^2$ Velorum, we adopt the FUV fluxes directly measured in \cite{gamma2Vel78}. 

In one important case - the star $\iota$ Orionis - the distance listed in \cite{McCallum2025} is known to be incorrect. The Hipparcos distance of $\sim 600$ pc listed for some stars in the Orion cluster is also inaccurate, and we therefore adopt a distance of 380 pc distance for those stars based on other measurements in the literature \citet{Sandstrom2007, Oplistilova2023}.

%%%%%%%%%%%%%%%%%%%%%%%%%%%%%%%%%%%%%%%%%%%%%%%%%%%%%%%%%%%%%%%%%%%%%%%%%%%%%%%%
\subsection{Dust Input to Radiative Transfer}
\label{subsec:technical}
%%%%%%%%%%%%%%%%%%%%%%%%%%%%%%%%%%%%%%%%%%%%%%%%%%%%%%%%%%%%%%%%%%%%%%%%%%%%%%%%

\subsubsection{Gridding}
\label{subsubsec:gridding}
%%%%%%%%%%%%%%%%%%%%%%%%%%%%%%%%%%%%%%%%%%%%%%%%%%%%%%%%%%%%%%%%%%%%%%%%%%%%%%%%

To determine the radiation field in 3D, we place representations of these stars inside the \citet{edenhofer2024} dust cube and apply radiative transfer techniques using the code \texttt{DIRTY} \citep{Gordon2001,Misselt2001}. The \citet{edenhofer2024} dust cube was constructed using the distances to stars measured with \textit{Gaia} parallax along with their separately measured extinction through the \textit{Gaia} BP/RP spectra. These data feed into an information field theory approach which returns a 3D dust map. This map is provided in arbitrary method-specific units that can be converted to standard dust opacity, though they do not contain any qualitative dust information (e.g. $R_V$). 

First, we construct a hierarchical grid. The \citet{edenhofer2024} map uses 516 logarithmically spaced distance bins spanning 69 pc to 1.25 kpc, with distance bin widths ranging from $\sim$0.4 pc near the Sun to $\sim$7 pc at the edge of the map. However, the effective spatial resolution at any given position is poorly constrained, as it depends not only on this distance binning but also on the local stellar density used in the 3D reconstruction of the map. We therefore adopt a variable voxel size chosen to roughly align with the effective binning of the \citet{edenhofer2024} map at various distances, while retaining computational efficiency and compatibility with the existing DIRTY infrastructure. 

We break the volume (a cube spanning $-1.25 \leq X, Y, Z \leq 1.25$ kpc) into ``megavoxels" which are 128 pc on a side and fully cover the volume. Megavoxels whose centers are within 200 pc of the Sun have a voxel size of 1 pc (128 $\times$ 128 $\times$ 128 voxels). Those between 200 and 400 pc have 2 pc resolution (64 $\times$ 64 $\times$ 64 voxels), those between 400 and 800 pc have 4 pc resolution (32 $\times$ 32 $\times$ 32 voxels), and those beyond 800 pc have 8 pc resolution. For each of these voxels we sample the \citet{edenhofer2024} map at that location to determine the dust volume density. 

Because the \citet{edenhofer2024} map is a sphere (with a radius of 1.25 kpc) centered on the Sun, the corners of the cube have no data. To account for the corners, we infill the dust map with a lower fidelity version of the \citet{edenhofer2024} map that extends out to 2 kpc. We do this only to provide a reasonable approximation for the radiative transfer outside the nominal 1.25 kpc-spherical-radius volume. Additionally, we note that \citet{edenhofer2024} does not provide a 3D map of the dust volume density within 69 pc of the Sun, and those cells are considered empty.

To convert from the native units of the \citet{edenhofer2024} map --- an arbitrary differential reddening unit based on \citet{Zhang2023} --- we first convert to $A_V$/pc using the extinction curve provided by \cite{Zhang2023}. Since the \citet{Zhang2023} extinction curve does not extend to the shorter wavelengths of the TD1 bands, we then convert from $A_V$/pc to the TD1 band wavelengths (1565~\AA, 1965~\AA, 2365~\AA, and 2740~\AA) using the \cite{Gordon2023} extinction curve. Finally, we convert to optical depth to ultimately obtain representations of $\tau_{1565}$/pc, $\tau_{1965}$/pc, $\tau_{2365}$/pc, and $\tau_{2740}$/pc.

To account for typical stars beyond the \citet{edenhofer2024} cube, which can be especially important for regions far from the Galactic disk midplane, we enhance the \texttt{DIRTY} code to include an option for repeating boundary conditions. We implement these repeating boundaries along the $X$ and $Y$ axes, which extend from $X,Y = -1.25$ kpc to $X,Y = 1.25$ kpc, in Galactic coordinates centered on the Sun. Hence, a photon being traced by the radiative transfer code that leaves the cube by crossing the boundary at $X=1.25$ kpc will re-appear in the cube with the same velocity vector and $Y,Z$ coordinates at the opposite cube boundary, $X=-1.25$ kpc. There are no repeating boundary conditions along the $Z$ axis. This mimics an infinite Galactic plane, which approximates Galactic radiation entering the cube from stars outside the boundary which are not included in the model. This also necessarily limits the accuracy of our map near the edges of the cube, however. We do not add any component to account for metagalactic UV (see \S\ref{subsec:ISRFvariation}).

\subsubsection{Dust Scattering Properties}
\label{subsubsec:dustscatteringprops}
%%%%%%%%%%%%%%%%%%%%%%%%%%%%%%%%%%%%%%%%%%%%%%%%%%%%%%%%%%%%%%%%%%%%%%%%%%%%%%%%

To execute the radiative transfer calculation, we must adopt values for the dust albedo $a$ and scattering phase function asymmetry $g$. We calculate these values from known measurements the diffuse Galactic light in the Milky Way. For both properties, the measurements are taken from \citet{Gordon04Review} in combination with \citet{Sujatha07, Sujatha05, Shalima04, Sujatha10}. The values of these measurements are not always in agreement and have significant uncertainties. We use linear interpolation to obtain values for the four TD1 bands (1565 \AA, 1965 \AA, 2365 \AA\ and 2740 \AA). For the scattering phase function asymmetry $g$, we find a line with a very shallow slope. We find that $g = 0.63$ for 1565 \AA\ and 1965 \AA, and $g = 0.64$ for 2365 \AA\ and 2740 \AA. We do the same for the albedo, but include a Drude profile for the dip in the albedo around the 2175 \AA\ feature. This dip in the albedo is due to the fact that the 2175 \AA\ feature is a pure absorption feature \citep{Calzetti95}. The Drude profile we include has the same width as the 2175 \AA\ feature. This fit showed an albedo that has a positive slope, with albedo $a$ equal to 0.45 for 1565 \AA, 0.4 for 1965 \AA, 0.39 for 2365 \AA\ and 0.45 for 2740 \AA. We note that the non-trivial albedo of astrophysical dust in FUV wavelengths is the primary reason we elect to use radiative transfer to do our modeling, as it captures the effect of the scattering of light by dust grains.

%%%%%%%%%%%%%%%%%%%%%%%%%%%%%%%%%%%%%%%%%%%%%%%%%%%%%%%%%%%%%%%%%%%%%%%%%%%%%%%%
\subsection{Computing the FUV ISRF}
\label{sec:fuvisrf}
%%%%%%%%%%%%%%%%%%%%%%%%%%%%%%%%%%%%%%%%%%%%%%%%%%%%%%%%%%%%%%%%%%%%%%%%%%%%%%%%

The radiative transfer code generates one radiation density cube for each of the four TD1 band wavelengths.
However, the FUV ISRF is often measured from 912 \AA\ to 2000 \AA~  \cite[e.~g.,][]{Draine78} so we generate a combined cube for convenient comparison to these commonly cited measurements. The TD1 bands do not fully cover this range, and thus to properly compute an FUV ISRF we perform a least-squares fit of the four TD1 intensities at each pixel to the typical FUV ISRF SED computed in \citet{Mathis1983} (see their table A3), and then infer the integrated FUV ISRF energy density. We find that we have a typical signal-to-noise (SNR) ratio of 10 for each separate TD1 band and 18 for this combined UV ISRF. The SNR is much higher near bright stars (see Figure \ref{fig:SNR}).

\begin{figure*}%[h!]
    \includegraphics[width=\linewidth]{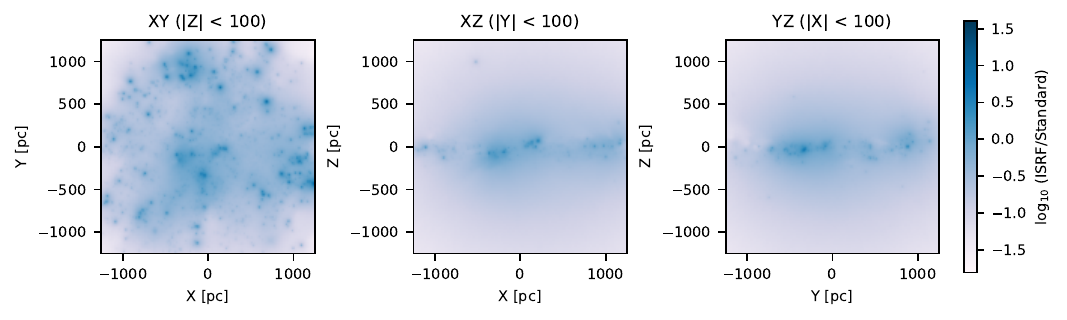}
    \caption{\textsc{LightCube slices.} Maps of the log of the mean radiation density of \textit{LightCube}, scaled to the Standard ISRF, from a top-down view of the disk (left panel) and vertical slices in $X$ and $Y$ (middle and right panel). Each slice is 200 pc thick and centered on the Sun.
    \label{fig:LCslices}}
\end{figure*}

\begin{figure}[b]
    \includegraphics[width=1.0\linewidth]{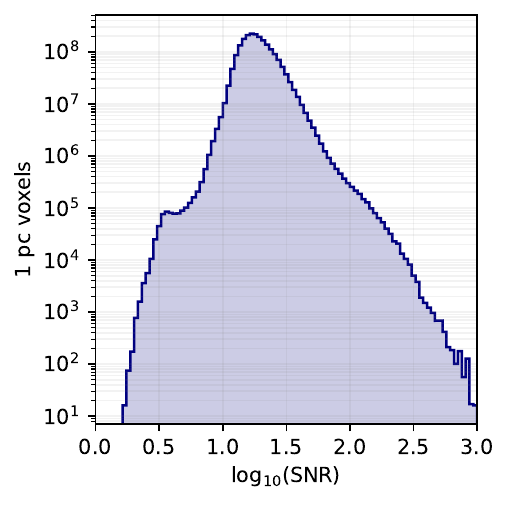}
    \caption{\textsc{SNR distribution.} The histogram of the signal-to-noise of the inferred \textit{LightCube} ISRF within 800 pc of the Sun.
    \label{fig:SNR}}
\end{figure}

%%%%%%%%%%%%%%%%%%%%%%%%%%%%%%%%% %%%%%%%%%%%% %%%%%%%%%%%%%%%%%%%%%%%%%%%%%%%%%
%%%%%%%%%%%%%%%%%%%%%%%%%%%%%%%%%%%% RESULTS %%%%%%%%%%%%%%%%%%%%%%%%%%%%%%%%%%%
%%%%%%%%%%%%%%%%%%%%%%%%%%%%%%%%% %%%%%%%%%%%% %%%%%%%%%%%%%%%%%%%%%%%%%%%%%%%%%
\section{Results}
\label{sec:results}

We provide the \textit{LightCube} final data products publicly to the community online\footnote{Data available for download at \url{https://hannahbish.me/LightCubes} and \url{https://app.globus.org/file-manager?origin_id=f2d103b7-ee4a-4f0b-bf1f-dc7e6df06b6f&origin_path=\%2F}}. For each of the four wavelengths, we provide an ISRF radiation energy density as well as an uncertainty on that value as separate data cubes. We also provide a combined UV ISRF energy density, scaled to the \cite{Draine78} Standard ISRF, with associated uncertainties. Figure \ref{fig:LCslices} shows slices of this UV ISRF. The typical signal-to-noise in the cube is 18, as shown in Figure \ref{fig:SNR}.

For convenience we provide the data on both a 1pc-resolution grid and a 4pc-resolution grid, since the full-resolution cube file is large. In order to map the values in the multi-resolution megavoxels onto the single-resolution cube, we simply average, replicate, or take the root-sum-square of the hierarchical DIRTY grid values in the megavoxels as is appropriate for the radiation densities and uncertainties in each region of \textit{LightCube}. We note that the uncertainties here are simple random uncertainties derived from the radiative transfer calculations; there are significant systematic uncertainties (discussed in \S \ref{sec:methods}) that should be taken into account when using these data products.

%%%%%%%%%%%%%%%%%%%%%%%%%%%%%%%%%%%%%%%%%%%%%%%%%%%%%%%%%%%%%%%%%%%%%%%%%%%%%%%%
\subsection{Local ISRF Intensity}
\label{subsec:ISRFintensity}
%%%%%%%%%%%%%%%%%%%%%%%%%%%%%%%%%%%%%%%%%%%%%%%%%%%%%%%%%%%%%%%%%%%%%%%%%%%%%%%%

We compute the radiation density at the Sun by averaging all voxels within 25 pc of the DIRTY grid origin, and find values of 5.23, 3.41, 2.39, and 1.57 $\times 10^{-17}$ erg cm$^{-3}$ \AA$^{-1}$ for the 1565 \AA, 1965 \AA, 2365 \AA\ and 2740 \AA, TD1 bands respectively. At the Sun we find the ISRF is 5.65 $\times 10^{-14}$ erg cm$^{-3}$ from 912 \AA\ to 2000 \AA, approximately 1.08 G$_0$ or 63.3\% of the Standard \cite{Draine78} ISRF. 

%%%%%%%%%%%%%%%%%%%%%%%%%%%%%%%%%%%%%%%%%%%%%%%%%%%%%%%%%%%%%%%%%%%%%%%%%%%%%%%%
\subsection{Completeness}
\label{subsec:completeness}
%%%%%%%%%%%%%%%%%%%%%%%%%%%%%%%%%%%%%%%%%%%%%%%%%%%%%%%%%%%%%%%%%%%%%%%%%%%%%%%%

\begin{figure}%[h!]
    \includegraphics[width=1.0\linewidth]{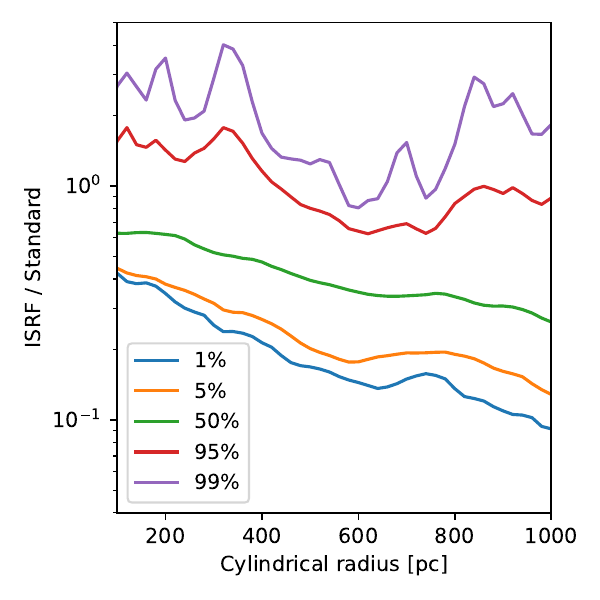}
    \caption{\textsc{\textit{LightCube} is complete out to 800 pc.} Percentile measurements of the ISRF in \textit{LightCube} as a function of cylindrical distance in the plane of the Galaxy within  \mbox{$|z| < $ 100 pc}. Incompleteness at the lower-luminosity end of the stellar sample at larger radii will cause the 1st and 5th percentile lines to drop off more quickly than the 99th and 95th percentile lines. We see this begin to occur at a distance of roughly 800 pc.
    \label{fig:percrad}}
\end{figure}

We find that within 800 pc of the Sun there is a detectable dropoff in ISRF intensity as a function of distance from the Sun, for both the highest and the lowest percentiles of the ISRF (see Figure \ref{fig:percrad}). This effect could be either be a decrease in the actual volume density of photons emitted from UV-bright stars, or a selection effect due to our input stellar data set's limited sensitivity (Malmquist bias). Because young stars are so heavily clustered, it is not obvious that we can rule out the first possibility. We believe our data set is quite complete out to these distances for the most luminous stars (see Figure \ref{fig:completeness}), and thus the drop off in the 99th and 95th percentile of the ISRF is likely due to the physical distribution of the most luminous stars. This can be seen visually on the map in Figure \ref{fig:edenhofer-map}, where there is a morphological drop off in the density of the most luminous stars beyond the solar neighborhood, as well as in Figure \ref{fig:completeness}, where there is a clear gap in bright stars between 400 and 800 pc from the sun. We believe the signature of incompleteness would show up as a steeper drop in the ``darkest" parts of the map (1st and 5th percentile), where the data set loses the least luminous contributing stars that fill in darker regions due to their much greater numbers. We see a \textit{modest} dropoff in the 50th, 5th, and 1st percentiles occurring past about 800 pc, which is consistent with the distribution seen in Figure \ref{fig:completeness}, and the location at which we begin to lose stars that dominate their 1 pc voxel. We therefore conclude that \textit{LightCube} is largely reliable out to at least 800 pc. The relative ``darkening'' of regions far from bright stars is modest beyond 800 pc, and thus \textit{LightCube} is likely still a valuable resource out to the boundaries of the \cite{edenhofer2024} volume at 1.25 kpc.

%%%%%%%%%%%%%%%%%%%%%%%%%%%%%%%%%%%%%%%%%%%%%%%%%%%%%%%%%%%%%%%%%%%%%%%%%%%%%%%%
\subsection{ISRF Variation}
\label{subsec:ISRFvariation}
%%%%%%%%%%%%%%%%%%%%%%%%%%%%%%%%%%%%%%%%%%%%%%%%%%%%%%%%%%%%%%%%%%%%%%%%%%%%%%%%

\begin{figure}%[h!]
    \includegraphics[width=1.0\linewidth]{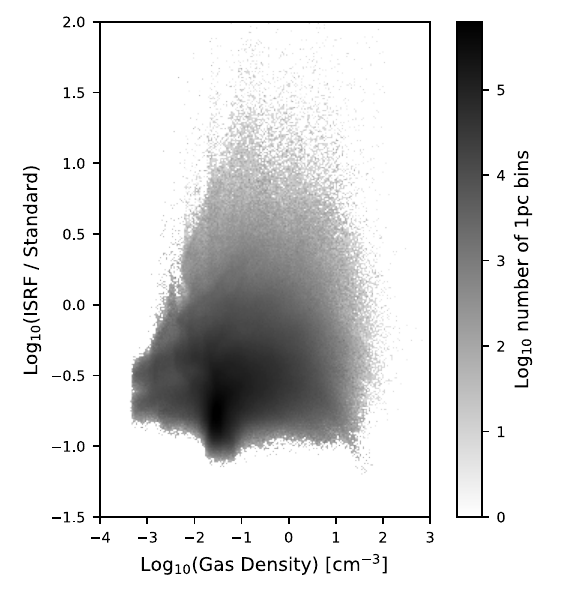}
    \caption{\textsc{ISRF and gas density.} Pixels in \textit{LightCube} within 800 pc of the Sun shown as a function of gas density and ISRF. Gas density is inferred from the \cite{edenhofer2024} based on standard Galactic dust-to-gas ratios. While a slight trend of less ISRF in denser regions is visible, very little of the overall variation in ISRF can be attributed to gas density.
    \label{fig:dustvrad}}
\end{figure}

Within 800 pc the radiation density of the ISRF is quite variable in the \textit{LightCube} model.  This variation is shown as a function of ISM gas density in Figure \ref{fig:dustvrad}. We compute gas density (in hydrogen atoms cm$^{-3}$) from dust density by assuming a standard Galactic reddening-to-hydrogen column density ratio reported in \cite{Bohlin78}. This works out to a conversion factor of 812 cm$^{-3}$/($A_V$/pc). We find that the variability of the ISRF is not primarily due to ISM density, but rather due to the distribution of stars. The 5\% to 95\% interval of \textit{LightCube} covers more than half an order of magnitude of variation. In higher density regions ($n > 30$ cm$^{-3}$) we find that ISRF varies by more than an order of magnitude over the 5\% to 95\% interval. 

In \textit{LightCube} the ISRF drops off vertically roughly as an exponential, with a scale height of 400 pc, and thus at 800 pc above the disk the ISRF is typically 12\% of the Standard ISRF. \cite{UVExGal} measure a metagalactic UV field of to $2\text{--}2.6\ \text{nW}\ \text{m}^{-2}\ \text{sr}^{-1}$, or 0.015 the Standard ISRF. As noted in \S \ref{sec:methods}, we do not include any metagalactic UV field in our model. This measured metagalactic field is more than order of magnitude lower than the \textit{LightCube} model 800 pc above the disk (and more than 0.5 dex lower even at the very top of \textit{LightCube}), and thus we suggest it can reasonably be ignored in most situations.

\begin{figure}%[h!]
    \includegraphics[width=1.0\linewidth]{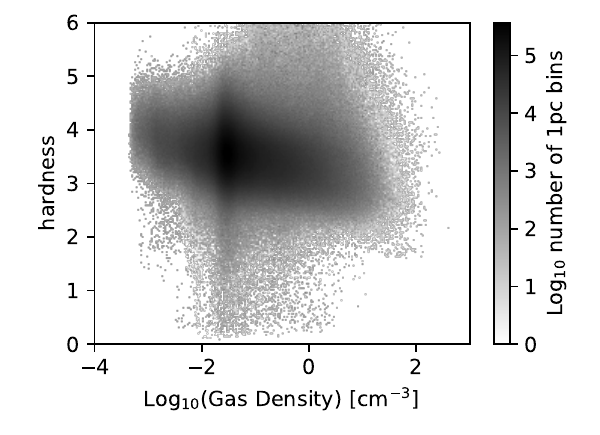}
    \caption{\textsc{Radiation field hardness.} The inferred volume density of the gas vs the hardness of the radiation field we measure with 800 pc of the Sun. The hardness shown here is the ratio of the 1565 \AA~ field to the 2740 \AA~field, and we only show voxels with 800 pc of the Sun.
    \label{fig:radvsdust-hardness}}
\end{figure}

%%%%%%%%%%%%%%%%%%%%%%%%%%%%%%%%%%%%%%%%%%%%%%%%%%%%%%%%%%%%%%%%%%%%%%%%%%%%%%%%
\subsection{Hardness}
\label{subsec:hardness}
%%%%%%%%%%%%%%%%%%%%%%%%%%%%%%%%%%%%%%%%%%%%%%%%%%%%%%%%%%%%%%%%%%%%%%%%%%%%%%%%

The ratio of the radiation field in our shortest wavelength band (1665\AA) to our longest wavelength band (2740 \AA) is a measure of the ISRF radiation ``hardness", which we show in Figure \ref{fig:radvsdust-hardness}. We see that there is some modest variation in hardness, and that there is a clear trend toward a softer UV field at higher volume densities, as would be expected given the higher extinction in the shorter wavebands.

%%%%%%%%%%%%%%%%%%%%%%%%%%%%%%%%%%%%%%%%%%%%%%%%%%%%%%%%%%%%%%%%%%%%%%%%%%%%%%%%
\subsection{Science Verification: the ISRF -- $R_V$ correlation}
\label{subsec:uvrv}
%%%%%%%%%%%%%%%%%%%%%%%%%%%%%%%%%%%%%%%%%%%%%%%%%%%%%%%%%%%%%%%%%%%%%%%%%%%%%%%%

\begin{figure*}[tp]
    \includegraphics[width=1.0\linewidth]{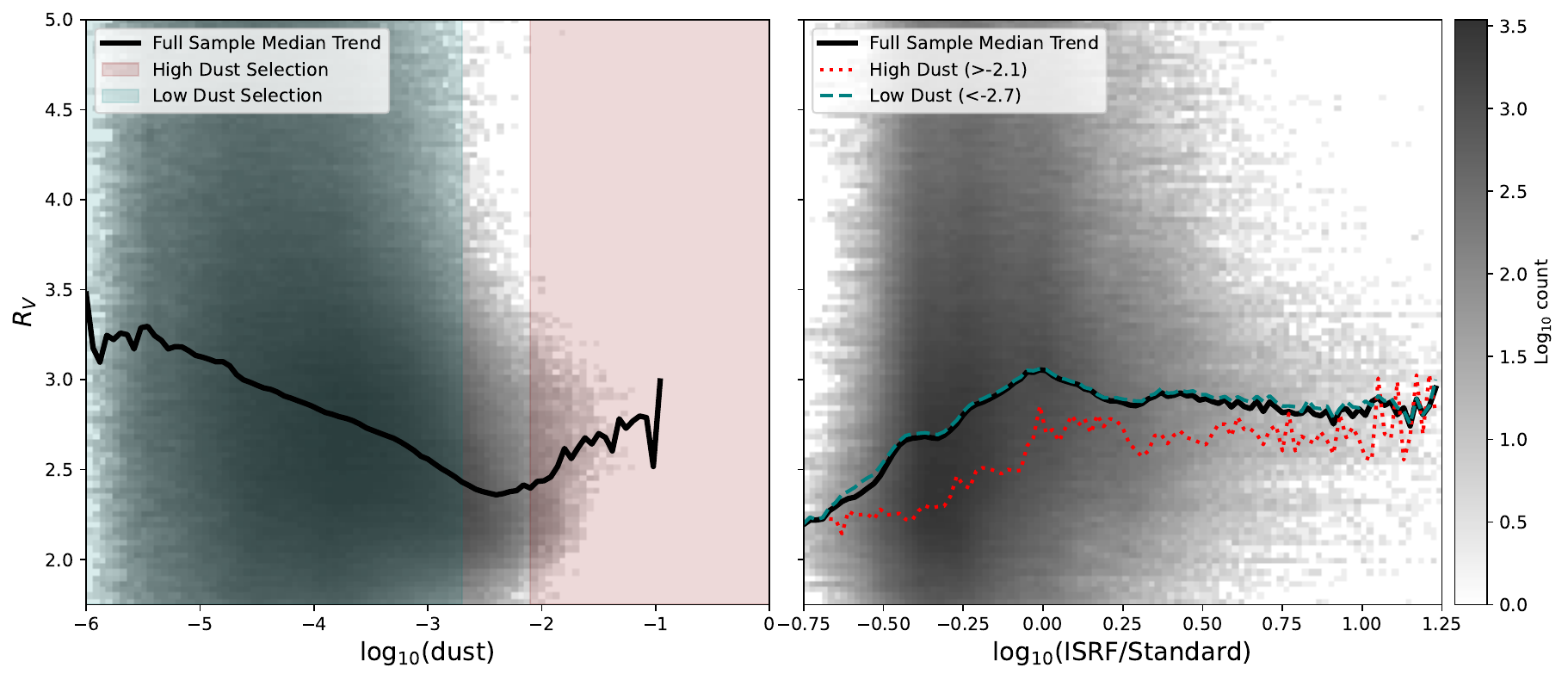}
    \caption{\textsc{ISRF-$\rm R_V$ correlation.} The dust -- $R_V$ (left) and ISRF -- $R_V$ (right)  relationship in the solar neighborhood. Both plots show points sampled every 2 pc in the volume $-500~\rm pc < X, Y < 500~pc $ and $-200~\rm pc < Z < -200~ pc $. In the left panel, the median $R_V$ measurement is shown with a black line. The `low density' region, where \cite{ZhangGreen25} finds a negative correlation between $R_V$ and dust density, is shaded blue, and the `high density' region, where they find a positive correlation, is shaded red. In the right panel, the same $R_V$ measurements are shown as a function of the ISRF in \textit{LightCube}. The ISRF-$R_V$ correlation is shown as a median trend with a black line. The blue and red lines represent the median trend for the corresponding blue and red shaded regions in the left panel. 
    \label{fig:uvrv}}
\end{figure*}

Recent work focusing on the diffuse ISM has shown that $R_V$ varies in unexpected ways \citep{Schlafly16}, including the surprising result from \cite{ZhangGreen25} that the ratio of total to selective extinction ($R_V$) in the solar vicinity initially decreases as dust volume density increases, then increases again, as shown in the left panel of Figure \ref{fig:uvrv}. Investigating the correlation of $R_V$ with radiation field density may provide insight into these variations. We compare the 3D distribution of $R_V$ from \cite{ZhangGreen25} to the ISRF measured in \textit{LightCube} by sampling every 2 pc in the volume $-500~\rm pc < X, Y < 500~pc $ and $-200~\rm pc < Z < -200~ pc $. The right panel of Figure \ref{fig:uvrv} shows the relationship between the \textit{LightCube} ISRF and the measured $R_V$ for these points. We find that there is a positive correlation between $R_V$ and radiation density for ISRF values below Standard. Above these values, the relation flattens and $R_V$ is no longer sensitive to the radiation field. In both dust density regimes, $R_V$ increases with ISRF at radiation densities below the Standard ISRF, then plateaus at higher radiation densities. This implies that the result is not simply a correlation to one of the trends shown in \cite{ZhangGreen25}. This result is presented for regions within 200 pc of the Galactic midplane where the signal is clearest, but it remains robust across a range of selection criteria.

%%%%%%%%%%%%%%%%%%%%%%%%%%%%%%%%% %%%%%%%%%%%% %%%%%%%%%%%%%%%%%%%%%%%%%%%%%%%%%
%%%%%%%%%%%%%%%%%%%%%%%%%%%%%%%%%% DISCUSSION %%%%%%%%%%%%%%%%%%%%%%%%%%%%%%%%%%
%%%%%%%%%%%%%%%%%%%%%%%%%%%%%%%%% %%%%%%%%%%%% %%%%%%%%%%%%%%%%%%%%%%%%%%%%%%%%%
\section{Discussion}
\label{sec:discussion}

%%%%%%%%%%%%%%%%%%%%%%%%%%%%%%%%%%%%%%%%%%%%%%%%%%%%%%%%%%%%%%%%%%%%%%%%%%%%%%%%
\subsection{Comparison to local ISRF Values}
\label{subsec:localISRFvalues}
%%%%%%%%%%%%%%%%%%%%%%%%%%%%%%%%%%%%%%%%%%%%%%%%%%%%%%%%%%%%%%%%%%%%%%%%%%%%%%%%

\citet{Bianchi2024} represents the most recent and complete work on the ISRF. This work finds an \textit{overall} higher value for the ISRF than had been found from previous works, and that the ISRF is somewhat redder. Their inferred ISRF from 13.6 to 6 eV is 4.13 10$^{14}$ erg cm$^{-3}$, significantly lower than the Standard ISRF and close to $G_0$. The \citet{Bianchi2024} result for the local FUV ISRF is very close to the \textit{LightCube} value of 5.65 $\times 10^{-14}$ erg cm$^{-3}$. As the \citet{Bianchi2024} value is derived almost entirely from optical and near-IR spectroscopy, and the \textit{LightCube} value is derived almost entirely from FUV photometry, the consistency of these results provides some additional confidence in both.

Additionally, \citet{Parravano2003} simulated the FUV radiation field in a region of the Galaxy with properties similar to the Solar circle, and examined the time-dependence of the field. They find that the FUV radiation field density undergoes significant fluctuations over approximately half an order of magnitude on timescales of a few hundred Myr, with less frequent stronger fluctuations across approximately 1.5 orders of magnitude on longer timescales of about 1 Gyr (see their Figure 4). They also find that the mean radiation field value is almost twice the median value due to these infrequent spikes in the radiation field, and that a significant fraction of the FUV field is dominated by a single source. This description of the variation of the FUV radiation field over time largely matches the spatial variation of the radiation field we see in \emph{LightCube}. 

%%%%%%%%%%%%%%%%%%%%%%%%%%%%%%%%%%%%%%%%%%%%%%%%%%%%%%%%%%%%%%%%%%%%%%%%%%%%%%%%
\subsection{Implications of ISRF variation in diffuse ISM}
\label{subsec:implications-diffuseISM}
%%%%%%%%%%%%%%%%%%%%%%%%%%%%%%%%%%%%%%%%%%%%%%%%%%%%%%%%%%%%%%%%%%%%%%%%%%%%%%%%

The ISRF variation in this map demonstrates that a constant UV ISRF in the local solar neighborhood, standardized by the seminal estimates of \citet{Habing1968} and \citet{Draine78}, is a poor approximation of true ISM conditions. Specifically, we find that the radiation field varies by more than 0.5 dex over the 5th-95th percentile in low density regions (0.01 cm$^{-3}~ < n < 10 \,\rm cm^{-3}$), and by more than order of magnitude in high-density regions ($n~>~30~\,\rm~cm^{-3}$).

The canonical two-phase description model of the neutral ISM --- in which a cold neutral medium (CNM; $T \sim 50$--$200$~K, $n \sim 30$--$100$~cm$^{-3}$) and a warm neutral medium (WNM; $T \sim 6000$--$10^4$~K, $n \sim 0.3$--$1$~cm$^{-3}$) coexist in thermal pressure
equilibrium --- rests fundamentally on a fixed heating rate dominated by photoelectric emission from small dust grains and polycyclic aromatic hydrocarbons \citep{Wolfire1995, Wolfire2003}. Because the photoelectric heating rate scales directly with the ISRF, the range of pressures over which a stable two-phase solution exists is itself a function of the local radiation environment.  A 0.5~dex variation in UV ISRF therefore implies a corresponding shift in the two-phase stability boundary along individual sightlines, such that CNM gas that is thermally stable in a low-field environment may find itself in the thermally unstable regime -- or even forced entirely into a WNM phase -- in a higher-field environment at the same total pressure. Observational studies that infer CNM and WNM fractions from 21~cm emission and absorption data \citep[e.g.,][]{Heiles2003,Murray2018} implicitly assume that the thermal equilibrium curve is universal. Our result suggests that some fraction of the scatter in observed spin temperatures, CNM column densities, and pressure measurements across sightlines may reflect intrinsic variation in the ISRF rather than just density or turbulent structure alone.

Furthermore, the abundance of molecular hydrogen (H$_2$) in the diffuse ISM is set by a balance between formation on grain surfaces and photodissociation, with the transition from atomic to
molecular gas occurring at a column density that depends sensitively on both the ISRF and the self-shielding of H$_2$ \citep{vanDishoeck1988}. A factor-of-three range in the ambient FUV field shifts the effective H\textsc{i}/H$_2$ transition threshold along individual sightlines, producing intrinsic scatter in observed $N({\rm H_2})/N({\rm H\textsc{i}})$ ratios that is not attributable to density structure. In addition, because CO requires shielding to survive photodissociation, the CO-emitting volume fraction of a cloud is strongly suppressed even at moderate increases in ISRF relative to the standard field. The CO-to-H$2$ conversion factor $X_{\rm CO}$ is thus sensitive to the incident radiation field precisely through this mechanism \citep{Wolfire2010, Bolatto2013}. A spatially varying ISRF at the 0.5~dex level implies that $X_{\rm CO}$ is not a fixed property of the diffuse ISM even at uniform metallicity and surface density, and that a significant and variable fraction of the molecular gas along any given sightline may be CO-dark \citep{Wolfire2010}.

%%%%%%%%%%%%%%%%%%%%%%%%%%%%%%%%%%%%%%%%%%%%%%%%%%%%%%%%%%%%%%%%%%%%%%%%%%%%%%%%
\subsection{Implications of ISRF variation in dense ISM}
\label{subsec:implications-denseISM}
%%%%%%%%%%%%%%%%%%%%%%%%%%%%%%%%%%%%%%%%%%%%%%%%%%%%%%%%%%%%%%%%%%%%%%%%%%%%%%%%

\begin{figure*}%[tp]
    \includegraphics[width=1.0\linewidth]{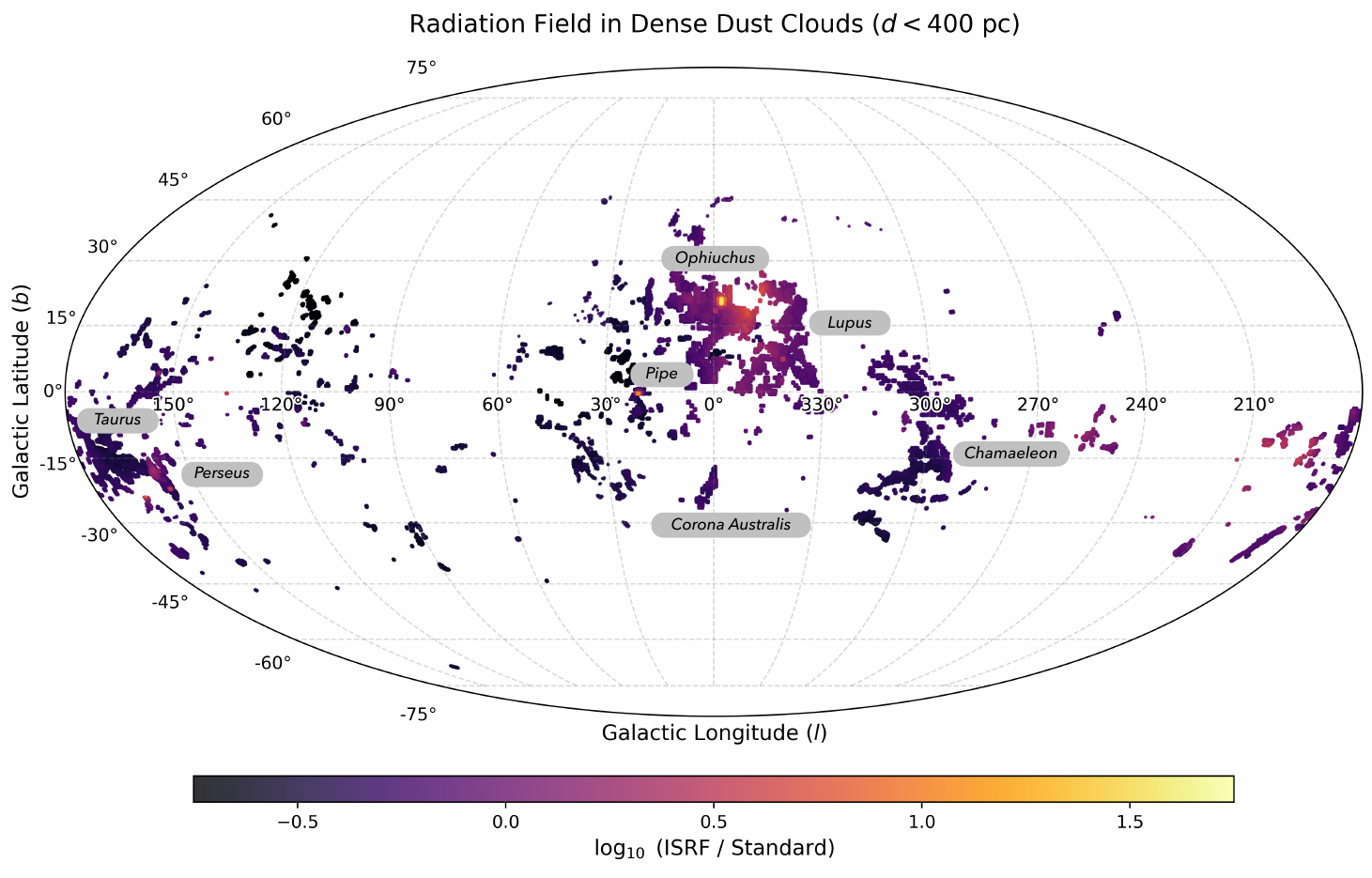}
    \caption{\textsc{The radiation field in dense clouds.} A projection of $n > 30$ cm$^{-3}$ gas on the Galactic sky towards molecular clouds within 400 pc of the Sun, including Taurus, Perseus, and the clouds associated with the Sco-Cen complex (Lupus, Chamaeleon, Corona Australis, Ophiuchus, and Pipe). Color indicates the \textit{LightCube} radiation field.
    \label{fig:moll}}
\end{figure*}

The ISRF varies by more than an order of magnitude over the 5th-95th percentile ranges in denser regions ($n > 30$ cm$^{-3}$). This large variation is expected; the most luminous UV sources are massive stars, and their proximity to nearby dense gas creates strong spatial gradients in the radiation field. \textit{LightCube} allows us to quantify this variation in 3D for the first time. 

Our volume includes dozens of well-studied molecular clouds. In Figure \ref{fig:moll}, we show the variation of the ISRF towards molecular clouds within 400 pc of the Sun, including Taurus, Perseus, and the clouds associated with the Sco-Cen complex (Lupus, Chamaeleon, Corona Australis, Ophiuchus, and Pipe). The contrast between Sco-Cen and clouds towards the Galactic anti-center (e.g. Perseus and Taurus) is striking; for example, when averaging over the entire $n > 30$ cm$^{-3}$ cloud volume, Ophiuchus (in Sco-Cen) experiences a typical ISRF roughly $3\times$ higher than Taurus. Sco-Cen has produced multiple generations of massive stars over tens of megayears \citep{Ratzenbock_2023} and the UV radiation from earlier generations is clearly illuminating the dense gas in which subsequent generations are forming. 

This difference has implications for the thermal, chemical, and star-forming conditions in molecular clouds. \citet{Rathjen_2025} use MHD simulations with a self-consistent, spatially varying FUV field to show that while non-ionizing UV radiation is subdominant to ionizing radiation, stellar winds, and supernovae in regulating the integrated star formation rate, it has subtle effects on the burstiness of star formation, likely through enhanced photoelectric heating on dust grains. More importantly, a realistic, spatially varying FUV field promotes a cold, diffuse molecular gas phase that is largely absent when a spatially constant, static ISRF is assumed. The wide-spread ISRF variations we see therefore imply that the abundance and distribution of this diffuse molecular component will vary substantially across nearby star-forming regions. For the first time, \textit{LightCube} provides the cloud-specific radiation conditions needed to quantify these effects in future.

%%%%%%%%%%%%%%%%%%%%%%%%%%%%%%%%%%%%%%%%%%%%%%%%%%%%%%%%%%%%%%%%%%%%%%%%%%%%%%%%
\subsection{Implications of the ISRF -- $R_V$ correlation}
\label{subsec:implications-RV}
%%%%%%%%%%%%%%%%%%%%%%%%%%%%%%%%%%%%%%%%%%%%%%%%%%%%%%%%%%%%%%%%%%%%%%%%%%%%%%%%

The general interpretation of $R_V$ variation is a shift in the average grain size: lower $R_V$ corresponds to smaller grains and higher $R_V$ to larger grains. This is supported by the finding that extinction curve shapes from the UV through NIR are strongly correlated \citep{Cardelli89, Gordon2023}. High $R_V$ curves are significantly ``grayer'' in the UV, a signature of grain growth. Because these correlations extend into the NIR—wavelengths sensitive to larger grains—variations are likely due to the evolution of the entire grain population rather than just a deficit of small grains. The positive correlation we find between the ISRF and $R_V$ at low radiation densities is unexpected; one might anticipate that the average grain size would decrease (lowering $R_V$) with an increasing radiation field due to radiation-induced grain destruction or shattering. 
%The flattening of $R_V$ near the Milky Way average of 3.1 at higher ISRFs suggests that significant grain evolution is restricted to low-radiation environments. 
One intriguing possibility is that the $R_V$ variation is driven by changes in the broad 2175~\AA\ bump. While \cite{ZhangGreen25} attribute $R_V$ variation to both accretion and coagulation, \cite{ZHG25} argue the effect stems specifically from the growth of carbonaceous nanograins through gas-phase accretion, as measured by the $q_{PAH}$ index. Since these nanograins are the favored carriers for the 2175~\AA\ bump, the red wing of this feature could drive $R_V$ variation. However, the measured bump strength variations in \citet{Gordon2023} ($A_\mathrm{bump}/A_V$ = 1.11 to 0.85 for $R_V$ = 2.2 to 5.6) are smaller than the factor of 2 suggested by \citet{ZHG25} as necessary to explain the $R_V$ variations reported by \citet{ZhangGreen25}. The flattening of the $R_V$ relation at radiation densities above the Standard ISRF suggests that whatever process drives this correlation -- whether or not it is nanograin destruction -- is inoperative beyond the Standard ISRF intensity. This result provides a significant constraint to any model of ISRF-driven $R_V$ variation.

%%%%%%%%%%%%%%%%%%%%%%%%%%%%%%%%%%%%%%%%%%%%%%%%%%%%%%%%%%%%%%%%%%%%%%%%%%%%%%%%
\subsection{ISRF Model Comparison}
\label{subsec:ISRFmodelcomparison}
%%%%%%%%%%%%%%%%%%%%%%%%%%%%%%%%%%%%%%%%%%%%%%%%%%%%%%%%%%%%%%%%%%%%%%%%%%%%%%%%

In parallel work, Lomeli et al. (in prep) have constructed an ISRF model using a complementary approach. Their work models the stellar UV fluxes from stellar atmosphere models and accounts for dust extinction, but doesn’t model the scattered radiation field.  
In contrast, \emph{LightCube} is constructed using observed UV fluxes from the TD1 catalog, and the resulting radiation field is computed using a radiative transfer code that includes scattering.  

There are strengths and weaknesses to both approaches. 
The Lomeli et al. model is not restricted to the discrete wavebands observed with TD1, does not suffer from statistical noise due to propagating a finite number of photon packets, and draws from a larger stellar sample using modern catalogs that can include more faint stars and are less prone to systemic errors than TD1.
However, our radiative transfer modeling allows for direct calculation of photon scattering, which impacts the radiation field significantly in higher-density regions where star-formation occurs.  Furthermore, we use directly measured UV fluxes, and are therefore less subject to uncertainties in the stellar parameters.  
With these differing approaches, both models will contribute to our understanding of the local ISRF.

%%%%%%%%%%%%%%%%%%%%%%%%%%%%%%%%% %%%%%%%%%%%% %%%%%%%%%%%%%%%%%%%%%%%%%%%%%%%%%
%%%%%%%%%%%%%%%%%%%%%%%%%%%%%%%%%%%% SUMMARY %%%%%%%%%%%%%%%%%%%%%%%%%%%%%%%%%%%
%%%%%%%%%%%%%%%%%%%%%%%%%%%%%%%%% %%%%%%%%%%%% %%%%%%%%%%%%%%%%%%%%%%%%%%%%%%%%%
\section{Summary}
\label{sec:summary}

In this work we present \textit{LightCube}, a three-dimensional model of the local ISRF in the UV. Using UV flux measurements from the TD1 catalog and stellar distances from \emph{Gaia} and Hipparcos, we identify stars that we expect to impact the UV radiation density in the nearby Galaxy and place them within a three-dimensional map of dust within 1.25 kpc of the Sun (\S\ref{sec:methods}, Figures \ref{fig:edenhofer-map} \& \ref{fig:completeness}). We use \texttt{DIRTY}, a radiative transfer code, to model photons emitted from the UV sources as they move through the volume and are absorbed or scattered by the dust. The result is a model of the ultraviolet ISRF in each of the four TD1 bands (1565 \AA, 1965 \AA, 2365 \AA, and 2740 \AA) as well as a combined model for the wavelength range from 912 \AA\, to 2000 \AA, which we provide for comparison to other measurements of the UV ISRF (\S\ref{sec:results}, Figure \ref{fig:LCslices}). The data cube has a variable resolution ranging from 1 pc near the Sun to 8 pc at the edges of the cube, with a typical SNR of 18 (\S\ref{subsubsec:gridding}, Figure \ref{fig:SNR}). We estimate that the UV sources are complete out to a distance of $\sim 800$pc, although the cube is likely still a valuable reference out to the full radius of 1.25 kpc (\S\ref{subsec:completeness}, Figures \ref{fig:completeness} \& \ref{fig:percrad}). We make these data products available to the community for general use\footnote{\url{https://app.globus.org/file-manager?origin_id=f2d103b7-ee4a-4f0b-bf1f-dc7e6df06b6f}}.

Our findings and conclusions are as follows:
\begin{enumerate}
    \item We calculate the ISRF at the Sun to be $5.65\times10^{-14}$ erg cm$^{-3}$ from 912 \AA\, to 2000 \AA. This is approximately 1.08 $G_0$, 63.3\% of the standard \citet{Draine78} ISRF, or 137\% of the \citet{Bianchi2024} local ISRF. (\S\ref{subsec:ISRFintensity}; Figure \ref{fig:LCslices})
    \item The modeled ISRF is quite variable, with more than an order of magnitude variation seen in dense regions ($n > 30$ cm$^{-3}$), and about half that in lower density regions. This variation is primarily due to the distribution of stars rather than the dust. The radiation energy density drops off vertically roughly as an exponential, with a scale height of 400 pc. The \textit{LightCube} model thus demonstrates that a constant UV ISRF in the local solar neighborhood, despite being a common assumption, is a poor approximation of true ISM conditions.(\S\ref{subsec:ISRFvariation} \& \S\ref{subsec:implications-diffuseISM}; Figure \ref{fig:dustvrad})
    \item There is modest variation in radiation hardness within the cube, with a trend toward a softer UV field at higher densities. (\S\ref{subsec:hardness}; Figure \ref{fig:radvsdust-hardness})
    \item By comparing LightCube to an estimate of the 3D distribution of total-to-selective extinction ratio $R_V$, we find a positive correlation between the radiation density and $R_V$ in regions with a radiation density below the Standard ISRF value. The relation flattens in regions with higher radiation densities. (\S\ref{subsec:uvrv} \& \S\ref{subsec:implications-RV}; Figure \ref{fig:uvrv})
    \item The potential applications of \textit{LightCube} are broad, and this data product is intended to be used as a tool by the astronomical community that enables further science. A detailed 3D map of the local ISRF can be used to investigate a range of topics including the conditions in molecular clouds (see \S\ref{subsec:implications-denseISM}) and photo-dissociation regions, thermal heating in the CNM, the impact of radiation fields on star formation, dust grain properties, the correlation of gas ionization state with local radiation field density, and more.
\end{enumerate}

There are a number of improvements that can be made to the \textit{LightCube} model in future. We are aware of a parallel paper (Lomeli et al. 2026, in prep) which presents a similar data product built using a somewhat different and complementary approach. We believe there are strengths and weaknesses to both approaches, and plan to combine these approaches to release a joint model (David Lomeli and Kedron Silsbee, private comm.).

%%%%%%%%%%%%%%%%%%%%%%%%%%%%%%%%% %%%%%%%%%%%% %%%%%%%%%%%%%%%%%%%%%%%%%%%%%%%%%
%%%%%%%%%%%%%%%%%%%%%%%%%%%%%%% ACKNOWLEDGEMENTS %%%%%%%%%%%%%%%%%%%%%%%%%%%%%%%
%%%%%%%%%%%%%%%%%%%%%%%%%%%%%%%%% %%%%%%%%%%%% %%%%%%%%%%%%%%%%%%%%%%%%%%%%%%%%%
\section{Acknowledgements}

The authors would like to express our most sincere gratitude to David Lomeli and Kedron Silsbee. When we discovered we were building similar data products, they were very gracious in working out an approach for simultaneous submission. The authors are extremely appreciative of their collegiality and cooperation.

The authors thank Alyssa Goodman and her group at Harvard for sharing their computing resources with us while we tested the earliest iterations of \textit{LightCube}.

The authors recognize the unceded ancestral land of the Piscataway and Susquehannock peoples, on which much of this research was conducted. We acknowledge the social, physical, spiritual, and kinship connections this land continues to share with Indigenous nations of the Susquehanna River and Chesapeake Bay. We recognize that we are uninvited visitors on Indigenous lands, and affirm that it is our responsibility to Indigenous nations to repair these unhealthy relationships and to steward all life.

This work was made possible in part due to NASA ADAP Grant 80NSSC22K0493, which provided support to H.V.B., J.E.G.P., S.E.C., and E.H.
C.Z. acknowledges support by the NSF CAREER award \#2442546. 

The authors acknowledge Interstellar Institute’s programs ``ii6” and ``ii7" and the Paris-Saclay University’s Institut Pascal for hosting discussions that nourished the development of the ideas behind this work. 

This work used the Anvil supercomputer at Purdue University through allocation PHY230181 from the Advanced Cyberinfrastructure Coordination Ecosystem: Services \& Support (ACCESS) program, which is supported by U.S. National Science Foundation grants \#2138259, \#2138286, \#2138307, \#2137603, and \#2138296.

This work has made use of data from the European Space Agency (ESA) mission {\it Gaia} (\url{https://www.cosmos.esa.int/gaia}), processed by the {\it Gaia}
Data Processing and Analysis Consortium (DPAC, \url{https://www.cosmos.esa.int/web/gaia/dpac/consortium}). Funding for the DPAC has been provided by national institutions, in particular the institutions participating in the {\it Gaia} Multilateral Agreement.

This work made use of the Hipparcos catalog from the European Space Agency (ESA). the SIMBAD database and VizieR service operated at the Centre de Données astronomiques de Strasbourg (CDS), NASA's Astrophysics Data System.

This research has made use of the SIMBAD database \citep{wenger2000} and the VizieR catalogue access tool \citep{Vizier2000}, CDS, Strasbourg Astronomical Observatory, France. 

This research has made use of the Astrophysics Data System, funded by NASA under Cooperative Agreement 80NSSC21M00561.

This research has made use of data provided by the High Energy Astrophysics Science Archive Research Center (HEASARC), which is a service of the Astrophysics Science Division at NASA/GSFC.

Software citation information was aggregated using \texttt{\href{https://www.tomwagg.com/software-citation-station/}{The Software Citation Station}} \citep{software-citation-station-paper,software-citation-station-zenodo}.

\software{
\texttt{astropy} \citep{astropy:2013, astropy:2018, astropy:2022},
\texttt{astroquery} \citep{astroquery2019, astroquery_17163526},
\texttt{DIRTY} \citep{Gordon2001,Misselt2001},
\texttt{extinction} \citep{extinction_barbary_2016},
\texttt{dust\_extinction} \citep{dust_extinction_18510679,dustextinction2024},
\texttt{dustmaps} \citep{dustmaps2018},
\texttt{Glue} \citep{glue_2015,glue_robitaille_2017},
\texttt{Matplotlib} \citep{Hunter:2007},
\texttt{NumPy} \citep{numpy},
\texttt{Python} \citep{python},
\texttt{scikit-learn} \citep{scikit-learn,sklearn_api,scikit-learn_17880109},
\texttt{SciPy} \citep{2020SciPy-NMeth, scipy_18736568},
\texttt{synphot} \citep{synphot2018}
}

%%%%%%%%%%%%%%%%%%%%%%%%%%%%%%%%% %%%%%%%%%%%% %%%%%%%%%%%%%%%%%%%%%%%%%%%%%%%%%
%%%%%%%%%%%%%%%%%%%%%%%%%%%%%%%%%%% APPENDIX %%%%%%%%%%%%%%%%%%%%%%%%%%%%%%%%%%%
%%%%%%%%%%%%%%%%%%%%%%%%%%%%%%%%% %%%%%%%%%%%% %%%%%%%%%%%%%%%%%%%%%%%%%%%%%%%%%
\appendix

\section{TD1 Data Quality}
\label{app:dataquality}
The TD1 Stellar Ultraviolet Fluxes Catalog \citep{Thompson1978,TD11995}, despite being the most recent all-sky catalog of UV-bright sources, was created in 1978. As a consequence there are a number of errors in the catalog which it is no longer feasible to adequately explain or correct for, since the astronomers involved in the project who might have the required information to do so are no longer active. While assembling and cross-matching the list of stars used to build LightCube, we discovered systematic errors in the coordinates of sources in the TD1 catalog. We note them here in the hope that it will be useful to others who wish to use these data. 

\begin{figure}[h!]
    \begin{center}
  \includegraphics[width=0.49\linewidth]{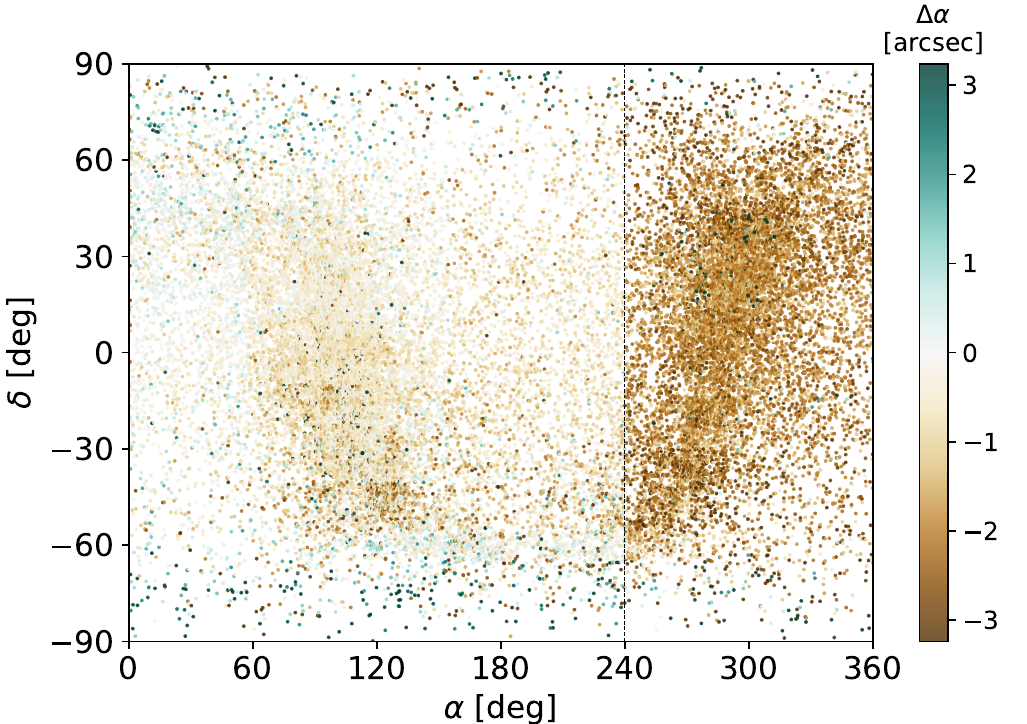}
  \hspace{-0.5em}\includegraphics[width=0.49\linewidth]{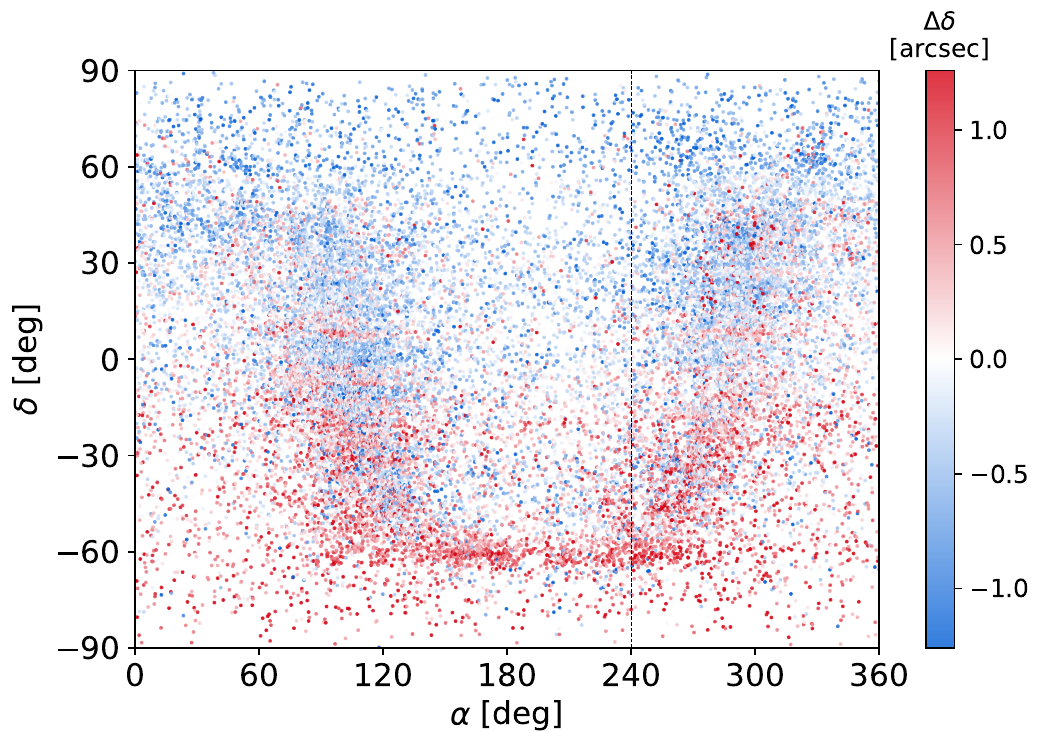}
  \caption{\textsc{Systematic errors in TD1 coordinates.} Discrepancy between right ascension (left panel) and declination (right panel) coordinates in TD1 and Simbad for stars matched on HD number. Positive values correspond to larger values for the TD1 coordinate, and negative values correspond to larger values for the Simbad coordinate. Note the large systematic error in RA at RA$>$240$^\circ$ (marked by the dashed line), and the systematic error in declination for positive versus negative values.}
  \label{fig:coordinate-offset}
      \end{center}
\end{figure}

The most glaring data quality issue was the accuracy of stellar coordinates. Figure \ref{fig:coordinate-offset} shows the systematic offset of the TD1 coordinates from the corresponding coordinates for each object matched by HD number in Simbad, for RA (top panel) and Dec (bottom panel). The TD1 RA coordinate offsets are relatively small from $0^\circ<\alpha<240^\circ$, but there is a discontinuity at $\alpha=240^\circ$, above which the discrepancies are much larger - often as much as 3 arcseconds. For declination, the TD1 coordinates are biased low at positive declination and biased high at negative declination, generally by approximately 1 arcsecond. We also note some striping in these declination error values where they fluctuate between positive and negative values.

Additionally, all stars in the TD1 catalog with no match in Simbad have TD1 ID numbers $>$30852 (with 31215 being the highest possible ID number). There are 244 stars in the catalog for which we found no match in either Gaia or Hipparcos, 223 of which have no Simbad match either. Most of these are fainter stars.
Furthermore, 728 sources in TD1 have no listed UV flux. Most of them have V-band magnitudes $>$ 9, but some are as bright as 2.2 in V-band.

\section{Optimizing Photon Packet Allocation for Radiative Transfer}
\label{app:photonalloc}
In the radiative transfer code DIRTY, each star is allocated a user-defined number of numerical photons (which we will call ``photon packets'' for the remainder of this section to distinguish them from the real photons emanating from stars). Since the brighter stars contribute the most to the ISRF, it is therefore computationally most efficient to allocate more photon tracers to the most luminous stars, ensuring their radiation fields have the highest signal-to-noise. Optimizing efficiency in allocating photons to the stars is important because running the radiative transfer code is computationally expensive, and computation time scales linearly with number of photons. Our goal is to reach a signal-to-noise of at least ${\rm SNR}_{targ}$ throughout the cube. We have investigated the most efficient way to allocate photons to ensure we are making the best use of the computing resources available to us, which we describe in the remainder of this section.

The number of packets to allocate to each star depends on each star’s luminosity $L_\star$, the resolution of the voxels, the background radiation level $U_0$, and the desired minimum SNR in the cube SNR$_{targ}$. $U_0$ is a first approximation to how much radiation all \textit{other} stars provide when calculating how many packets to allocate to an additional star, which we adopt from existing measurements of the UV ISRF.

\subsection{Setting up the problem}

We aim to allocate $\gamma_{j,{tot}}$ packets to each star $\star_j$. These packets allow the radiative transfer code to estimate the radiation field $\langle S_i \rangle$ in each of the voxels $V_i$. The radiative transfer code returns an estimate for the radiation field from each star provided to each voxel $\langle S_{i,j}\rangle$ as well as an estimate of the error in that radiation field $\langle \sigma_{ij}\rangle$. We target a signal to noise ratio in each $V_i$ of at least ${\rm SNR}_{targ}$

\subsection{Initial allocation}

To establish an initial allocation of $\gamma_{1,j}$ packets to each star $\star_j$, we assume a starting background radiation field $G_0$ that exists in all cells with zero error. Each star has a luminosity $L_j$, and contributes a radiation field at $V_i$ of
\begin{equation}
    G_{ij} = \frac{L_j}{4 \pi c d_{ij}^2}
\end{equation}
where $d_{ij}$ is the distance between $\star_j$ and $V_i$ and $c$ is the speed of light. The error in this radiation field is
\begin{equation}\label{eqn:scale}
    \sigma_{ij} = c_{\rm MC}\gamma_{1,j}^{-1/2}\frac{L_j}{4 \pi c d_{ij}^2}
\end{equation}
where $c_{\rm MC}$ is a scaling constant that we can determine by doing test runs of DIRTY code. We find the signal to noise ratio of 
\begin{equation}\label{eqn:minsnr}
    SNR_{ij} =\frac{\frac{L_j}{4 \pi c d_{ij}^2} + G_{0}}{\sigma_{ij}}
\end{equation}
is at a \textit{minimum} when 
\begin{equation}\label{eqn:disteq}
    \frac{L_j}{4 \pi c d_{ij}^2} = G_{0}
\end{equation}
(i.e., the stellar radiation is equal to the level of the background radiation field).

Thus
\begin{equation}
    {\rm SNR}_{min} = \frac{2}{c_{\rm MC}\gamma_j^{-1/2}}
\end{equation}
and the optimal photon allocation is therefore
\begin{equation}
    \gamma_j = \left(\frac{{\rm SNR}_{min} c_{\rm MC}}{2}\right)^2
\end{equation}

$c_{\rm MC}$ can be determined from the output of a test run.

In practice, this approach must be modified for a hierarchical grid, in which cells vary in size. 

\subsection{Hierarchical grid considerations}

Through numerical experiments, we find that while the noise in a cell does scale as expected in terms of distance from the star and number of packets assigned to the star (Equation \ref{eqn:scale}), it does not scale as we would naively expect with cell size ($\sigma \propto \Delta l^{3/2}$). Instead, we find that:
\begin{equation}
    \sigma_{ij} \propto \Delta\ell_{i}^{1.15}
\end{equation}
where $\Delta \ell_{i}$ is the side length of cell. This is not terribly surprising, given the complexity of how packets are handled in Monte Carlo radiative transfer codes like DIRTY. To determine the optimal allocation of packets to each star we use the following procedure. 
First, we first find the distance at which the stellar radiation field equals the assumed background field (Equation \ref{eqn:disteq}) and use the results of our numerical experiments to find the number of packets needed to reach our $SNR_{targ}$ in the smallest grid ($\Delta\ell_{0}$) that we have in the hierarchical DIRTY grid at that distance from the star:
    \begin{equation}\label{eqn:packets}
        N_{packets,\,U_\star=G_0} = 2.11 \times SNR_{targ}^2 \times \left(\frac{d}{pc}\right)^2 \times \left(\frac{\Delta\ell_0}{pc}\right)^{-2.3}
    \end{equation}
Second, we find both the nearest and the farthest points from the star for each grid resolution finer than $\Delta\ell_{0}$, and determine the number of packets required to reach $SNR_{targ}$ for each of those locations:
    \begin{equation}\label{eqn:nearfar}
        N_{packets, near/far} = 2.11 \times \left(\frac{L_\star}{4 \pi d^2 c U_0 + L_\star}\right) \times SNR_{targ}^2 \times \left(\frac{d}{pc}\right)^2 \times \left(\frac{\Delta\ell}{pc}\right)^{-2.3}
    \end{equation}
Lastly, we choose the maximum of the packets we calculated in Equations \ref{eqn:packets} and \ref{eqn:nearfar} and use it as our final packet allocation for the star.

In practice this results in a number of packets to use for each star and each TD1 filter, which depends on the luminosity of the star, the resolution of the voxels near the star's location, and the target signal-to-noise. For reference, in the 1565 \AA~band we allocate 5,012,357 packets to our most luminous star, which also has the highest packet allocation.

\bibliography{LightCube, JP_refs}{}
\bibliographystyle{aasjournalv7}

\end{document}